\documentclass[preprint,onecolumn,a4paper,showpacs,amsmath,amssymb,endfloats]{revtex4}
\usepackage{graphicx}

\usepackage{dcolumn}
\usepackage{bm}
\usepackage{color}

\usepackage{mathptmx}

\newcounter{saveeqn}

\begin{document}
\title{Influence of structural disorder and large-scale geometric fluctuations on the Coherent Transport of Metallic Junctions and Molecular Wires}
\author{R. Maul and W. Wenzel}
\affiliation{Forschungszentrum Karlsruhe, Institut f\"ur
  Nanotechnologie, Postfach 3640, 76021 Karlsruhe, Germany} 
\affiliation{Institut f\"ur Theoretische Festk\"orperphysik
and DFG-Center for Functional Nanostructures (CFN),
Universit\"at Karlsruhe, 76128 Karlsruhe, Germany}
\date{\today}
\begin{abstract}
Structural disorder is present in almost all experimental measurements of electronic transport through single molecules or molecular wires. To assess its influence on the conductance is computationally demanding, because a large number of conformations must be considered. Here we analyze an approximate recursive layer Green function approach for the ballistic transport through quasi one-dimensional nano-junctions. We find a rapid convergence of the method with its control parameter, the layer thickness, and good agreement with existing experimental and theoretical data. Because the computational effort rises only linearly with system size, this method permits treatment of very large systems.
We investigate the conductance of gold- and silver wires of different sizes and conformations. For weak electrode disorder and imperfect coupling between electrode and wire we find conductance variations of approximately 20\%. Overall we find the conductance of silver junctions well described by the immediate vicinity of  narrowest point in the junction, a result that may explain the observation of well-conserved conductance plateaus in recent experiments on silver junctions. 
In an application to flexible oligophene wires, we find that strongly distorted conformations that are sterically forbidden at zero temperature, contribute significantly to the observed average zero-bias conductance of the molecular wire. 
\end{abstract}
\maketitle
\section{Introduction}
The study of electron transport properties at the molecular or even atomic scale has generated many striking insights in the last decade~\cite{nitzan:2003,kwiatkowski:2008,heurich:2002,oetzel:2008}. Many features of the measured transport characteristics of molecular wires in breakjunctions or AFM/STM setups~\cite{reed:1997,cui:2001} could be explained on the basis of the Landauer formalism. There has been tremendous progress in the  characterization of metallic nano-wires \cite{scheer:1997,scheer:1998,agrait:2003,thygesen:2003,xie:2004}, organic nano-wires \cite{tao:2006,choi:2008} and nanotubes \cite{tans:1997,postma:2001,mceuen:2004}. Unfortunately, due to the complexity of the experimental setup, most of these measurements have no control over the details of the electrode geometry, which results in a spectrum of IV-characteristics. 
In the face of these and other uncertainties theoretical models can help understand and explain the experimental data \cite{rodrigues:2002}). These calculations are challenging because the electronic structure of the central cluster of the junction (,,extended molecule''), comprising the molecular wire and some part of the electrode fragments, must be computed in atomistic models. Because of the high cost of accurate electronic structure calculations, IV-characteristics are often computed for idealized geometries and electrode configurations \cite{strange:2008}. 

Due to computational limitations most theoretical studies of molecular junctions have employed idealized junction geometries, including placement and contact of the molecule with respect to the electrode, as well as the electrode geometry. Many present day fabrication techniques for nanojunctions entail strong deformations of the system, e.g. rupture of electrode wires (in break-junctions) and subsequent contacting procedures, which make it unlikely that ideal junction geometries are ever realized in practice. Despite the use of accurate electronic structure methods, only semi-quantitative agreement with the measured conductance has been obtained in a number of studies to date. Imperfect electrode geometries and contacts are likely to contribute to this widely discussed discrepancy \cite{evers:2004}, but are difficult to assess experimentally. 

The effects of structural disorder, of impurities and non-ideal electrode geometries are difficult to asses with transport methods that relies on computationally demanding ab-initio electronic structure theory, which often scale as $O(N^3)$ with the system size. One possibility to take such effects into account is to develop efficient approximations\cite{andrews:2008}, which allow for inclusion of larger electrode fragments in electronic structure calculation of the ,,extended molecule'' region. While such approximate methods may not predict the absolute value of the conductance quantitatively, they may help explain differential features that may originate from the existence of a large structural ensemble in the junction geometry. Such methods are also useful to investigate the coherent transport properties in very large, complex molecules, for which high-accuracy calculations remain infeasible to date. 

In previous work tight binding models were successfully applied by Pauly~\textit{et al.} \cite{pauly:2006} to investigate conductance histograms and strain forces of Ag, Pt and ferromagnetic Ni break junctions. Additionally H\"afner~\textit{et al.} \cite{haefner:2008} studied the influence of the absence of magnetic domains in the ferromagnetic 3$d$ materials Fe, Co and Ni. In the case of metal-molecule-metal junctions based on oligophenylene wires the length-dependent conductance and thermopower as well as elastic and photo-assisted transport properties, a $\pi$-orbital tight binding model has given intriguing insights \cite{pauly:2008b, viljas:2008}. 

In this paper we investigate convergence of a recursive layer Green's function (RLGF) approach\cite{verges:1999} that scales linearly with the system size.  This approach permits the treatment of very large systems comprising more than a thousand atoms, e.g. DNA~\cite{zalinge:2006} or the atomic transistor \cite{xie:2008b}, as well as computations of ensembles of thousands of structural models for disordered systems. We first analyze the truncation effects on the accuracy of conductance calculations for a several systems, where much experimental and theoretical data is available. Gold \cite{novaes:2006,rego:2003} and silver \cite{pauly:2006,rodrigues:2002} molecular wires are among the most widely studied systems in molecular electronics. We find that even for metallic systems, where the electronic wave functions are most extended, the conductance converges rapidly with the truncation cutoff parameter and converges to values in good agreement with experiment and accurate prior studies~\cite{pauly:2008b,strange:2008}, at a fraction of their computational cost. 

Next we apply this method to novel, computationally challenging applications on transport through molecular wires. We investigated different scenarios for static and dynamic disorder in metal\cite{scheer:1997,scheer:1998,agrait:2003,thygesen:2003,xie:2004,novaes:2006,rego:2003,pauly:2006,rodrigues:2002}  and metal-organic molecular\cite{choi:2008} wires. 
In an application on the influence of a non-vanishing tilt- and twinning-angle between the source/drain electrode tips and find a variation of up to 20\% in the zero bias conductance. Additionally we probe the effect of randomly introduced surface vacancies on the electrode tips, which can lower the conductance by up to 30\%.

We then turn to silver nano-clusters, which are promising systems for applications as electronic materials and for surface nano-structuring. While their structural properties have been extensively investigated \cite{doye:1998, shao:2008, ledo:2007, ramnani:2007}, much less is known about their transport properties. We present the first theoretical study of electronic transport through large clusters (260 atoms) and study the dependence of the transport properties on the structure and size of the nano-cluster.

While these examples concern mononuclear systems, the presence of metal-organic junctions significantly complicates the physics of nanoscale transport. Charge transfer, differences in the accuracy of the description of electronic structure of the electrode and the organic molecule, as well as other factors, complicate the theoretical description of such systems \cite{heimel:2006,preuss:2005}. We have investigated the effect of dynamic disorder on the zero-bias conductance of oligophenylene molecular wires as a function of their length and geometry. We compare the total transmission function obtained by the layer approximation with the full device calculation and find a good quantitative agreement at a fraction of the computational cost.

Recent investigations have focused on the importance of even small scale geometric fluctuations on molecular transport. Here we extend this work to the first investigation of large-scale conformational change. We generate a large thermodynamically relevant ensemble of, which include complete ring-flips of the molecules, comprising thousands of conformations,  to  discuss the dependence of the conductance change on the torsion angle fluctuations. We find this system to be a striking example, where large scale structural fluctuations significantly affect the conductance. Configurations, which are sterically forbidden at low temperature, contribute significantly to the overall time-average of the zero-bias conductance.

\section{Methods}

\begin{figure}
\resizebox{0.45\textwidth}{!}{\rotatebox{0}{\includegraphics*{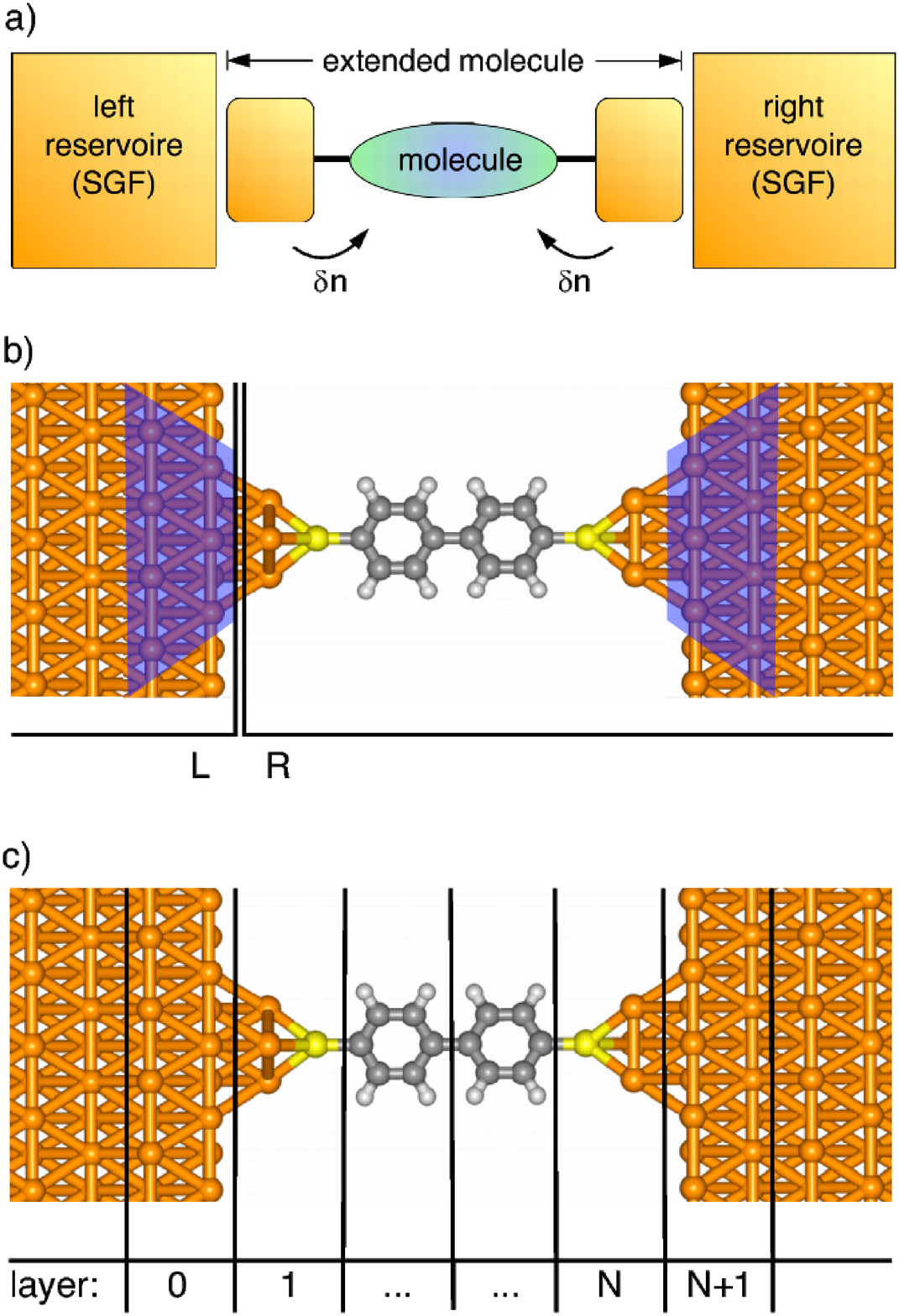}}} 
\caption{(Color online) Schematic representation of the different regions of a single molecule junction in the Landauer approach. (a) Definition of the extended molecule including a fraction of the electrodes. This permits the natural charge transfer $\delta n$ and screening effects as an organic system is attached to a metal surface. (b) Realization of a molecular junction by a phenyl-ring-based wire coupled via sulphur atoms to the [111]-layers of gold leads. The semi-infinite leads are represented by their surface Green's function defined on the atoms of the dark shaded area. Horizontal lines indicate a possible division into a left and and a right part of the system. (c) Representative division of the system into principal layers to illustrate the recursive Green's function approach.}
\label{fig:AuMolAu}
\end{figure}

The electronic conductance of nanoscale junctions is often analyzed in the framework of Landauer theory \cite{landauer:1970,buettiker:1985}, which divides the system conceptually into idealized electrodes and a central region called ,,extended molecule'', that contains the nanojunction of interest (see Figure~\ref{fig:AuMolAu}a). Using Landauer's formula, the current is computed as:
\begin{equation}
        I(V) = \frac{2e}{h} \int dE \tau(E,V) [f_L(E) - f_R(E)] 
\end{equation}
where $\tau$ denotes the total transmission function. In order to evaluate the transmission function of the extended molecule, we must compute its electronic structure in an adequate atomistic model \cite{damle:2002}. Equation (1) stipulates the existence of perfect Fermi-seas at the appropriate chemical potential of the electrodes. For these quantities to have meaning, electrode fragments must be included into of sufficient size into the extended molecule region, so that charge transfer and image-charge effects can be taken into account \cite{heimel:2006,preuss:2005}. The extended molecule for which the transmission must be computed may thus be much larger than the physical object which is placed between the electrodes. In particular for homonuclear systems, such as metallic wires, there is no clear-cut distinction between electrode and "system" at all. 
 
To compute the current, we consider the total charge of the system on the right side of an arbitrary division of the system into a right (R) and left (L) part, respectively (see Figure 1(b)):
\begin{equation}
\rho = -e \sum_{i \in \text{RHS}} c^{\dagger}_i c_i,
\end{equation}
where $c^{\dagger}_i$ creates an electron in orbital $i$ and the sum runs over all orbitals on the right hand side. 
The current $I$ can then be obtained in perturbation theory from the equation of motion $i\hbar I = [H, \rho] = -i e \hbar v_z$, where  
\begin{equation}
i \hbar v_z = \sum_{i \in L } \sum_{j \in R }  t_{ij} (c_i^{\dagger} c_j - c_j^{\dagger} c_i)
\end{equation}
defines the velocity operator \cite{verges:1999}. In the expression of the electron velocity operator the indices $i$ and $j$ run over the orbitals in the left (L) and right (R) parts of the system, respectively. $v_z$ is proportional to the transfer matrix element $t_{ij}$ times the electron propagation, expressed by the particle creation and annihilation operators $c^{\dagger}_i$ and $c_i$, respectively. Calculation of the charge flux at any arbitrarily chosen interface is thus sufficient to obtain the total current $I = -e \langle v_z \rangle$ through the system.  

Using Kubo's formula \cite{verges:1999,damle:2002} we can evaluate the coherent zero bias transmission $\tau$ at $T = 0$ as 
\begin{equation}
\tau = \text{Tr} \left[ (i \hbar v_z) \text{Im} \ \mathcal{G}(E) \ 
(i \hbar v_z) \text{Im} \ \mathcal{G}(E)\right].
\end{equation}
The imaginary part of the Green's function $\text{Im} \ \mathcal{G}(E)$ is related to the advanced/retarded  Green's functions $\mathcal{G}^a=[\mathcal{G}^r]^{\dagger}$ by
\begin{equation}
\text{Im} \ \mathcal{G}(E) = \frac{1}{2i} \left[ \mathcal{G}^r(E) -  \mathcal{G}^a(E) \right].
\label{eqn:img}
\end{equation}
The computation of $\mathcal{G}^r$ in general requires the inversion of the Hamilton operator of the extended molecule of the system. The computational effort of this calculation grows rapidly with the system size (i.e. the total number of orbitals), restricting the size of the system and the complexity of the electrode fragments that can be included in the extended molecule.  

For this reason, we divide the extended molecule into a set of principal layers perpendicular to the current flow direction and describe the electrons contained in one principle layer with a block Hamiltonian matrix. This introduces a truncation parameter which was found to be well converging with an increasing layer thickness (see results section A). We take only these overlap matrix elements into account, which belong to nearest neighboring layers and obtain a Hamiltonian matrix of band-diagonal type. The main advantage of this approach is that the computational effort of the matrix inversion scales linearly with the system size.  

We employ this feature in the evaluation of the transmission function using a Recursive Green's function scheme\cite{verges:1999}. This method is based on an iteration over the principal layers from the left end to the right end of the extended molecule. For every layer we calculate the Green's function by inverting the corresponding block Hamiltonian and take the influence of the semi infinite right region via self energies into account. The iteration ends up with the calculation of the Green's function of the left most principal layer, which is plugged together with the surface Green's function and the corresponding coupling matrixes into Kubos formula for conductance.

The whole method is based on the assumption of non-interacting electrons permitting the use of a tight binding like model Hamiltonian (extended H\"uckel), that allows for an efficient computation of the electronic structure during a dynamic process. More details of the numerical methods are described in the appendix.

\section{Results}

\subsection{Convergence of the layer approximation}

Metallic nanowires have been among the first and most widely studied systems in molecular electronics \cite{pauly:2006,novaes:2006,rego:2003,rodrigues:2002}. Metallic systems are often most challenging for linear-scaling electronic structure methods, because the electronic wave functions are extended. To provide a stringent test for our ,,local'' approximation, we have investigated the convergence of the layer approximation for two representative examples, namely gold- and silver-wires respectively. 

\begin{figure}[t]
\resizebox{0.45\textwidth}{!}{\rotatebox{0}{\includegraphics*{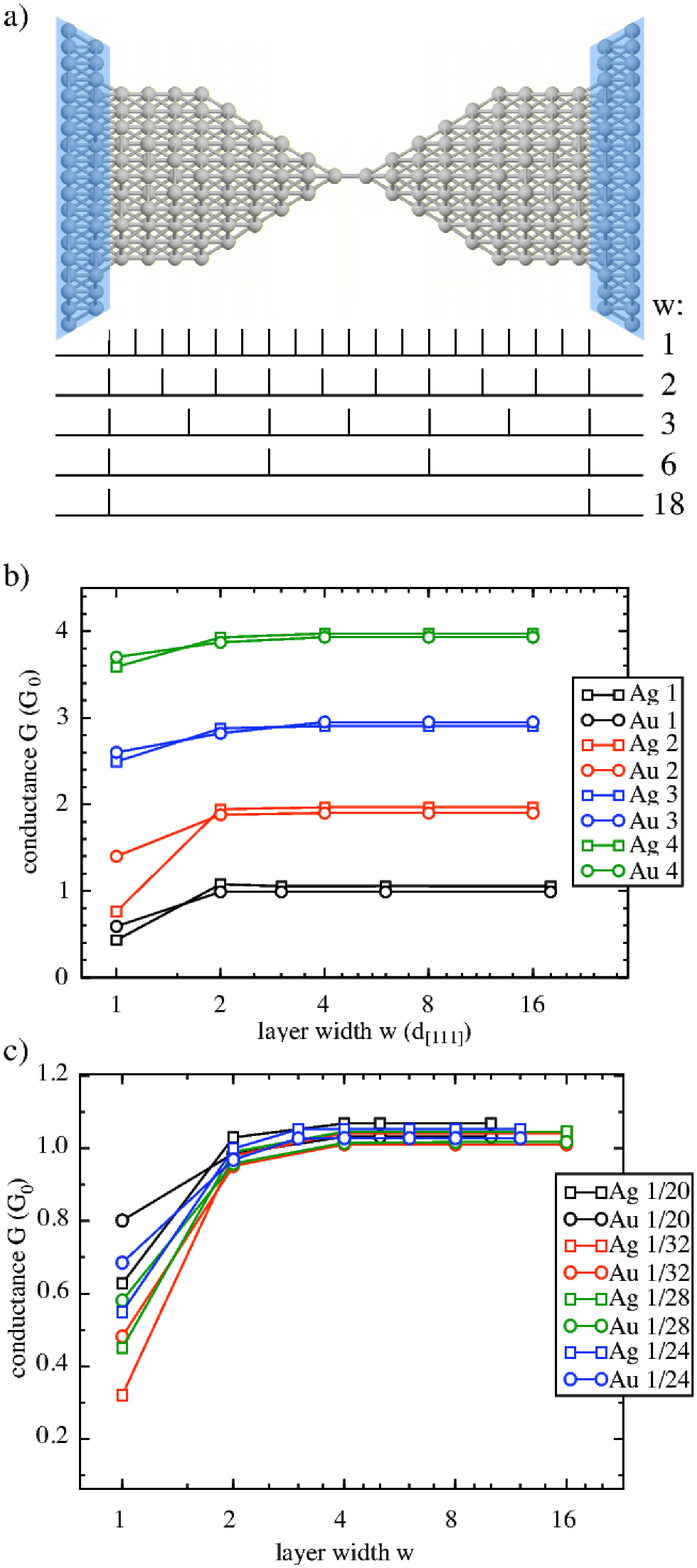}}} 
\caption{(Color online) Convergence test of the conductance depending on the ,,principal layer''-thickness for several nano-junctions. (a) Model nanojunction of 45.2 \AA{} length and minimal cross section of one atom allowing for several different principal layer divisions, indicated by the marked lines below the conformation. On the right hand side the thickness of the principal layers of the actual division is indicated, respectively. $w$ is given in units of the [111] atomic layer distance $d_{[111]} = 2.35$\AA{} (b) Corresponding conductance values for the upper described sets of ,,principal layers'' for a silver and a gold contact, respectively. The dependence of the conductance on the layer division is also shown for similar metallic junctions with a minimal cross section of 2, 3 and 4 atoms, respectively. (c) Metal quantum wires with one conductance quantum, but with increasing length between 20 and 32 atomic layers show the same rapid convergence behavior with increasing principal layer thickness. }
\label{fig:convtest}
\end{figure}

We begin the investigation by dividing silver and gold model junctions into a set of ,,principal layers'' with increasing thickness $w$. In order to establish the convergence of the method for large systems, the test geometry has to be of sufficient length. Here we investigate junctions of 45.2~\AA{} length in $z$-direction, containing 388 silver or gold atoms with a nearest neighbor distance of 2.88~\AA{} in both metals \cite{guang:2006,qian:2006}. We prepare the electrodes as perfect fcc-clusters, which narrow towards the center to form a single-atom point contact at their tips, generating a dimer structure which permits a current flow in the crystallographic [111] direction. The extended molecule region and the layer divisions are illustrated in Fig.~\ref{fig:convtest}a. The bulk electrodes are designated by the two larger layers on each side of the system. 

We calculate the conductance for varying widths $w = 1,\dots, 18$ d$_{[111]}$ of the ,,principal layers'' (Fig.~\ref{fig:convtest}a), i.e. the full length of the extended molecule region. Figure~\ref{fig:convtest}b shows that the conductance of the silver and gold model junction as a function of the principal layer thickness $w$ converges rapidly to the experimental value. The same holds true for the junction conformations labeled Ag 1 - Ag 4 and Au 1 - Au 4, which have a minimum cross section of 1-4 atoms, respectively. Furthermore, we calculated the convergence of the conductance for longer wire geometries (Fig.~\ref{fig:convtest}c), constructed by sequentially introducing additional atoms into the point contact at the minimum cross section. Thereby we obtain silver and gold junctions of 20, 24, 28 and 32 atomic layers in the $z$-direction.

\begin{figure}[t]
\onecolumngrid
\centering
\resizebox{0.75\textwidth}{!}{\rotatebox{0}{\includegraphics*{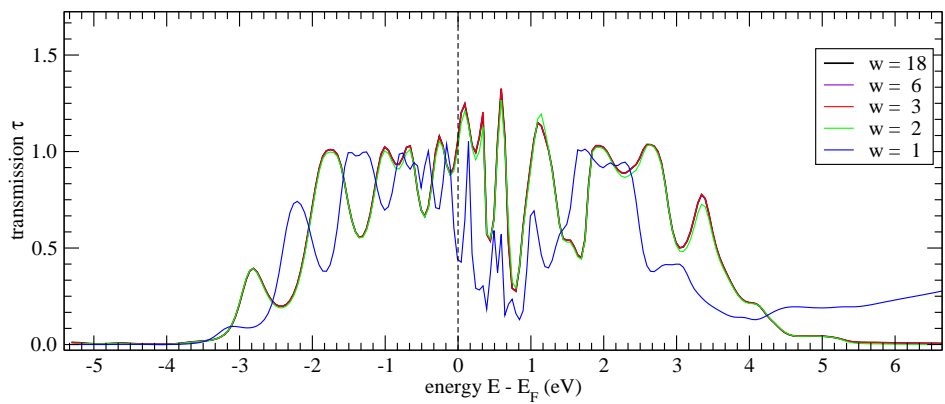}}} 
\caption{(Color online) The total transmission $\tau(E)$ of the model silver junction shown in Fig 1a, in the different levels of approximation, i.e. a changing number of ,,principal layers''. Convergence with respect to the principal layer thickness is achieved as soon as $w$ becomes is larger than 2.}
\label{fig:testsys5}
\end{figure}

For $w=1$ the conductance is significantly underestimated to approximately 0.5 G$_0$ for both metals, indicating that hopping processes across distances larger than the interatomic distance are important. For all choices of the layer thickness with $w > 1$ the conductance has converged to nearly the experimental values. For $w = 3$ the layer division retains the symmetry of the [111] crystal stacking order ''ABCABC...'' in fcc-latices. We investigate the convergence in more detail at the level of the transmission in Fig~\ref{fig:testsys5}, which shows the total transmission function $\tau(E)$ of the geometry shown in Fig.~\ref{fig:convtest}a over an energy interval $[E_F - 6 \text{ eV}, E_F + 6 \text{ eV}]$. Again we find that all curves for $w>2$ agree well with one-another.

In order to demonstrate the efficiency of this method we compare the computation time of the transmission curves shown in Fig 2. With a resolution of $\Delta E = 10$ meV the transmission of the system divided into 1, ..., 18 layers required 1368, 594, 429, 336, and 294 seconds, respectively. Using this approximation, e. g. with 6 principal layers, that takes 31\% of the time of the ,,full-device'' calculation, while increasing only a neglectable los of accuracy.

\subsection{Imperfect Electrode Geometries}

\begin{figure}
\resizebox{0.45\textwidth}{!}{\rotatebox{0}{\includegraphics*{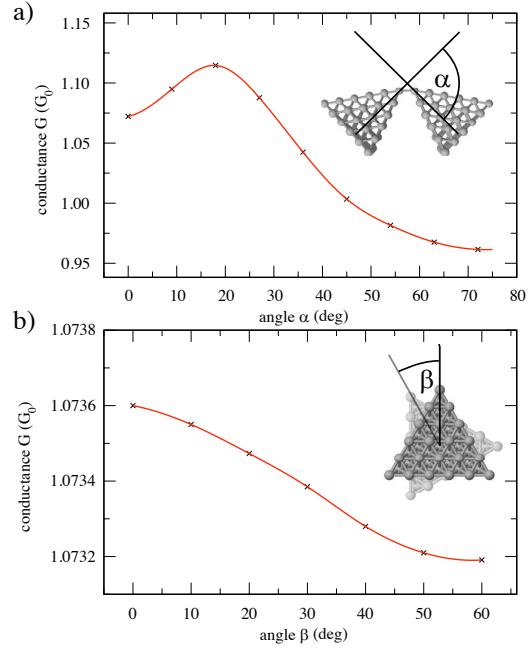}}} 
\caption{(Color online) The tilt- and twinning-angle dependence of the conductance of a silver point contact. (a) Only a moderate change of $\langle G \rangle$ is observed during tilting the electrodes up to 70 degrees. (b) The twinning of the electrodes between 0 and 60 degrees results in a nearly constant conductance.}
\label{fig:tiltangle}
\end{figure}

There are two obvious parameters which define the junction geometry with respect to the electrodes that are presently not under experimental control: the tilt and twinning angles of the two electrode fragments with respect to one another. In order to investigate the dependence of the conductance on these parameters, we have prepared a ideal fcc-silver junction with 224 atoms as in the previous section and varied the tilt ($\alpha = 0, ..., 70$ deg) and twinning ($\beta = 0, ..., 60$ deg) angle, as shown in the insets of Fig.~\ref{fig:tiltangle}. Increasing the tilt-angle $\alpha$ from 0 to 20 degrees leads to a slight increase of the conductance by circa 0.05 G$_0$ which can be explained by the influence of interference effects, which strongly depend on small changes of the atomic positions. A further increase of $\alpha$ from 20 to 70 degrees results in a decreasing conductance by 0.15 G$_0$, which corresponds to the loss of crystal symmetry across the junction. In contrast, twinning the electrode from 0 to 60 degrees leads only to a minor change in the conductance of $4\cdot 10^{-4}$~G$_0.$

When an electrode is manufactured in a break junction or generated by contacting the tip of an AFM/STM, it is very unlikely that the perfect lattice geometries with perfect surfaces along the crystalline axis, assumed in nearly all theoretical investigations, are realized in practice. According to all models of electronic transport, each surface defect creates an additional scattering center that may impede coherent transport through the junction. Imperfections in the geometry of the electrode tips will thus influence the ballistic transport. On the other hand, we have seen in the previous section, in agreement with many prior studies\cite{pauly:2006,novaes:2006,rego:2003,rodrigues:2002}, that the conductance of the junction is mostly determined by its most narrow region\cite{xie:2008a}. 

In order to estimate the significance of tip disorder we have therefore prepared a perfect junction as above and then randomly removed atoms from the surface of the electrode in the vicinity of the contact point. 
The number of silver atoms in the extended molecule region is systematically decreased by removing 28 atoms  at randomly chosen surface positions. To maintain coherent transport, the two central atoms were never removed. Every junction geometry with $n\cdot28$ vacancies $(n = 1, ...,5)$
was generated 500 times, with randomly chosen vacancy positions. For each conformation we computed and subsequently averaged the conductance. The calculations were performed using a principal layer thickness parameter $w=3$. 

\begin{figure}
\resizebox{0.45\textwidth}{!}{\rotatebox{0}{\includegraphics*{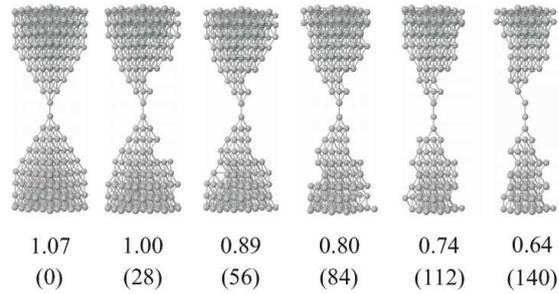}}} 
\caption{(Color online) Representative examples of the generated conformations of silver nanojunction with an increasing number of surface vacancies. The presence of defects leads to a decrease of the conductance by up to 30\%, indicated by the corresponding conductance values below the geometries. The number of vacancies in conformation is given in brackets.}
\label{fig:vacancies}
\end{figure}

\begin{figure}
\resizebox{0.35\textwidth}{!}{\rotatebox{0}{\includegraphics*{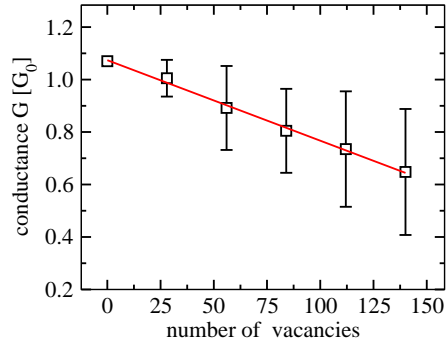}}} 
\caption{(Color online) Conductance values averaged over conformations with equal number of surface impurities. The error-bars indicates that the change in the conductance depends less on the number of defects, but more on their positions.}
\label{fig:vac_statistic}
\end{figure}

Samples of the resulting junction conformations are shown in Fig. \ref{fig:vacancies} with the corresponding averaged conductance value in units of G$_0$, respectively. In addition, the total number of surface vacancies on the current junction geometry is given below the conductance values. Figure~\ref{fig:vac_statistic} shows the resulting conductance values averaged over conformations with equal number of surface impurities. Creating 140 vacancies, which is half of the initial number of atoms, reduces the total conductance by 40\% in average. The rapidly increasing size of the error-bars indicates that the change in the conductance depends strongly on their positions.

\subsection{Silver Nanoclusters}

\begin{figure}
\resizebox{0.65\textwidth}{!}{\rotatebox{0}{\includegraphics*{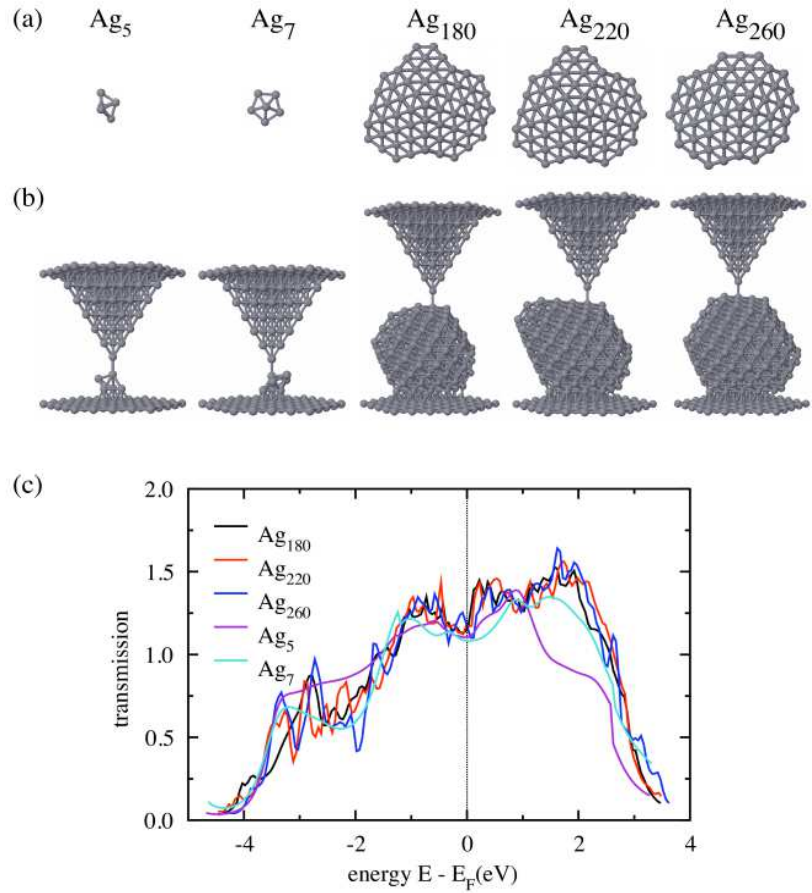}}} 
\caption{(Color online) Transmission of silver nano-clusters. (a) Shows the top view on the examined nano-cluster conformations with 5, 7, 180, 220, and 260 atoms. (b) Cluster conformations energetically optimized on a silver substrate layer with a pyramidal electrode on top. (c) Calculated transmission function of the junction conformations shown in "b". The vertical line indicates the Fermi energy. }
\label{fig:cluster}
\end{figure}

Recent experiments of silver junctions\cite{xie:2004, xie:2008a} suggest a strong stability of the observed zero-bias conductance in electrochemically grown silver junctions. While this effect was locally explained\cite{xie:2008b} by the selection of specific contact geometries, the overall shape of the silver contacts is likely to vary strongly from one realization of the next. To assess the effect of these large-scale geometric differences, we have prepared locally similar, but globally different junction geometries by placing silver clusters of various size in different orientations on a perfect surface and then contacting the tip of the cluster with an "ideal" junction. 
For this purpose we use the optimized cluster geometries from Ref.~\cite{shao:2008, doye:1998}, which where generated by Monte Carlo minimization and the modified dynamic lattice search method. Figure~\ref{fig:cluster}a shows the top view of the studied silver clusters with 5, 7, 180, 220, and 260 atoms and decahedron ($m$-Dh) core symmetry. As illustrated in Fig.~\ref{fig:cluster}b we consider the metal clusters attached to a silver substrate layer of the crystallographic [111] direction. The second electrode is realized by a pyramidal tip on top of the nano-cluster similar to a STM-setup. 

We optimized the position of the silver nano-cluster on the substrate using a Metropolis Monte-Carlo technique combined with the semi-empirical Gupta potential for the silver atoms as described above. During the simulation the silver cluster is treated as a rigid body, so only translations and rotations of the cluster are allowed - structural rearrangements insight the cluster are forbidden. The metal cluster surface consists of a set of [111], [110] and [100] facets. The minimum of the potential energy is reached, if the system is arranged such that the largest [111] facet (which is always the largest subsurface in the present cases) and the [111] substrate layer are facing each other. The top electrode is assumed to point directly on one arbitrarily chosen silver atom on the cluster surface. 

Figure~\ref{fig:cluster}c shows the total transmission function of the clusters Ag$_{5}$, ... , Ag$_{260}$. The conductance of the systems is given by the average of the transmission over a small interval around the Fermi energy $[E_f-\Delta, E_f+\Delta]$ with $\Delta = 50$ meV. For the clusters with 5, 7, 180, 220, and 260 atoms we find conductance values of 1.10, 1.08, 1.15, 1.17, and 1.17 G$_0$, respectively, which means, that the conductance is less effected by the size of the nano-cluster and depends more on the point contact to the second electrode. This observation may explain the observed stability of the experiment: While reconstruction of the junction geometry assures the selection of a specific local geometry, the overall conductance depends only very little on the global shape of the clusters forming the contact. This result is also in good agreement with an earlier study~\cite{bascones:1998} using a simplified model based on random matrix theory.

We also note that an irregular fluctuation of the transmission as a function of energy is observed, which increases with the cluster size. Such fluctuations can be conceptually explained by the 
 interference of the incident electron waves with waves scattered repeatedly in the extended molecule region containing the sliver nano-cluster and the electrode tip. Increasing the cluster size permits scattering processes of increasing order which leads to conductance fluctuations of higher frequency. An analysis of the average energy spacing of the extrema of the transmission (which may be measured by applying a gate voltage) can help to estimate the size of the backscattering region.

\subsection{Thermal fluctuations of the conductance of oligophenylene wires}

So far we have discussed only mononuclear extended molecule regions. It is well known that the presence of metal-organic interfaces complicates electronic structure and, as a result, electronic transport calculations. We have therefore studied the coherent conductance of phenyle-di-thiol (PDT), a ,,Drosophila'' of molecular electronics. Because this molecule was studied extensively in the past~\cite{strange:2008,evers:2004,kondo:2006,tomfohr:2004,emberly:2003,reed:1997, varga:2007,ventra:2000}, it allows for a comparison of the RLGF approach with experiment and various other levels of theory. We investigate the transmission of oligophenylene molecules of varying lengths, which lend themselves nicely to an investigation of the layer approximation in an organic, semi-conducting system. The structure of the molecules suggests a natural introduction of layers in terms of single phenyl-ring units, similar to the layers introduced by base-pairs in DNA\cite{starikov:2005}.

\begin{figure}
\resizebox{0.45\textwidth}{!}{\rotatebox{0}{\includegraphics*{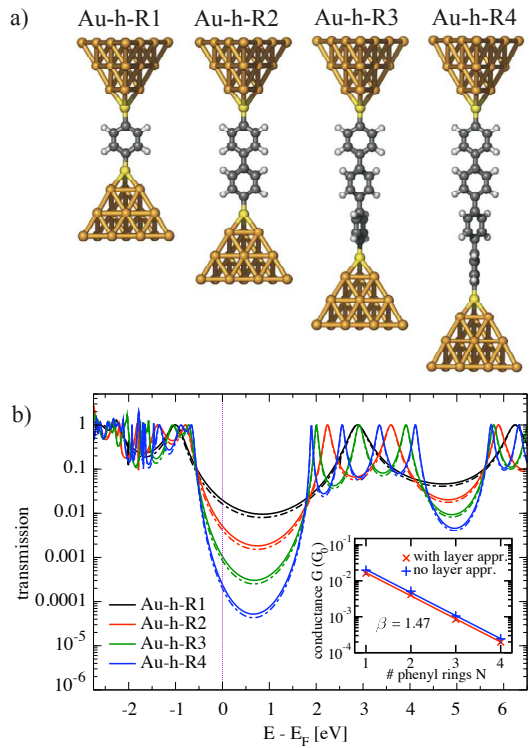}}} 
\caption{(Color online) Configurations of the organic molecular wires studied in this work. (a) Oligophenylene molecules covalently bond to Au$_{19}$ clusters along the cristallographic [111] axis. One phenylene ring unit represents one principal layer. (b) Total transmission as a function of the energy of the shown oligophenylens with (dashed line) and without (solid line) the principal layer approximation in good qualitative agreement with the DFT results of Ref. \cite{pauly:2008b}. In the layer approxiation one principal layer contains a single phenyle ring unit. The vertical line indicates the Fermi energy. (\textit{Inset}) Length dependence of the conductance of the oligophenylene wires. The conductance decreases exponentially with the number of the phenyle rings in the wire in good agreement with experimental data.}
\label{fig:molwires}
\end{figure}

Figure~\ref{fig:molwires}a illustrates the oligiphenylene molecules covalently bound to Au$_{19}$-clusters, using the same notation as for the conformations in Ref.~\cite{paulyPHD:2007}. The molecule is connected to the Au electrodes at both sides by a symmetric covalent bond of a sulfur atom to three Au atoms. In the literature this bonding situation is referred to as the hollow position \cite{paulyPHD:2007}. The electrode clusters where constructed from fcc lattices as above, while the geometry of the phenyl wires was optimized using density functional theory (DFT) in the local density approximation (LDA)~\cite{ahlrichs:1989,wohlthat:2007,paulyPHD:2007}. As can be seen in Fig.~\ref{fig:molwires}a there is a non-vanishing tilt-angle between the phenylene-rings, which varies between 33.7 and 34.5 degrees due to the interplay of  steric repulsion and $\pi$-conjugation of adjacent rings. A detailed investigation of the influence of (conjugation induced) tilting on the coherent transport properties in biphenyl-derived dithiols was recently given in Ref. \cite{pauly:2008a}.

In Fig.~\ref{fig:molwires}b the total transmission is shown as a function of energy for the molecular junctions above, once with (dashed line) and once without the layer approximation (solid line). Due to the neglect of several overlap matrix elements the transmission and conductance obtained with the layer approximation is below the full device transmission (see table~\ref{tab:opi}). 
The measured conductances of oligophenylene wires with amine end groups indicate even lower conductance values (also given in table~\ref{tab:opi}) which may arise from the differences in the coupling to the electrodes. 
With increasing length of the phenyl wire the transmission gap decreases from 3.89 eV to 2.56 eV. The equidistant transmission at the Fermi energy of the different molecular wires indicates the correct exponential decrease of the conductance with linear increasing wire length \cite{viljas:2008,wold:2002,wakamatsu:2006,venkataraman:2006}. 
The proportionality of the conductance decay $G/G_0 \sim e^{-\beta N}$ of the oligophenylene wires once with and once without the layer approximation is shown as inset in Fig.~\ref{fig:molwires}b. In both cases we obtain a decay coefficient $\beta = 1.47$  which is close to the experimental value $\beta_{exp.} = 1.5$ reported in \cite{venkataraman:2006} for amine end groups. Nevertheless, the comparison between theory and experiment remains difficult because of the different end groups used.

Next we investigate the influence of thermally induced molecular vibrations on the coherent transport properties of an Au-h-R4 wire. For the simulation of the dynamics of the system we use of the AMBER 8 molecular dynamics package \cite{amber8}, which employs the well established GAFF-forcefield and a Langevin thermostat method to model temperature. Assuming fixed gold atoms of the electrodes we simulate the evolution of the system at 300 K for 10 ps in time steps of 2 fs. Every 10-th time step a snapshot of the conformation is taken as input for the conductance calculation generating 500 conformations for analysis. For each conformation we calculate the zero-bias conductance. Within the simulation period we find repeated conductance fluctuations by more than an order of magnitude. 

\begin{figure}
\resizebox{0.85\textwidth}{!}{\rotatebox{0}{\includegraphics*{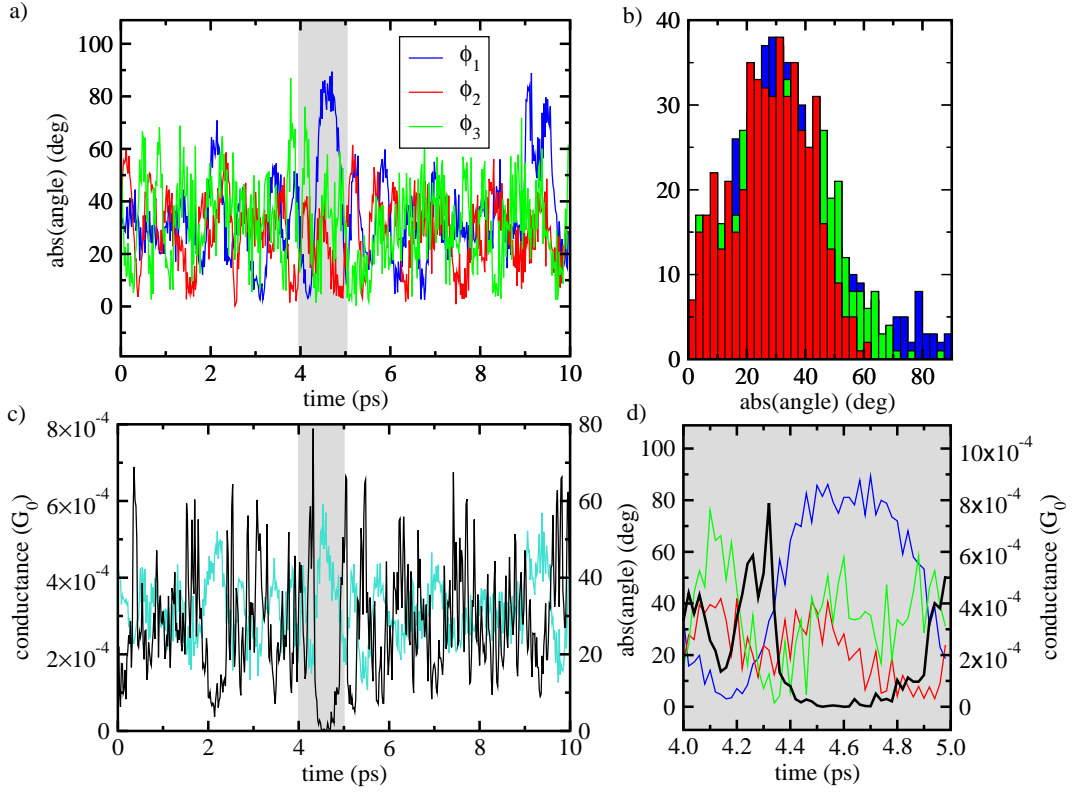}}} 
\caption{(Color online) Thermal influence on the conductance of a molecular wire at 300 K.(a) Fluctuation of the torsion angles $\phi_1$, $\phi_2$ and $\phi_3$ between the ring-units occuring in Au-h-R4, respectivlely. (b) Distribution histogram of the frequency of occurrence of a particular torsion angle. (c) Corresponding conductance (black) and average torsion angle $\bar{\phi}$ (gray/turquoise) at the fluctuation process during 10 ps simulation time. (d) Zoom into the 4ps-5ps range, which shows that a short-time increase of the torsion angles (thin gray/colored curves, left axis) leads to a strong decay of the total transmission (bold black curve, right axis) of the nano wire.}
\label{fig:fluct}
\end{figure}

Recent investigations have already focused on the influence of intramolecular vibrations on the conductance \cite{xie:2008b}. Here, we find  an interesting model system where thermal fluctuations induce large-scale conformational change. The conductance of a conformation correlates highly with its "planarity", because the fully planar conformation leads to a strong overlap of the $\pi$-orbitals, which in turn increases the transmission. However, such planar configurations are forbidden at zero temperature because of steric repulsion of the hydrogen atoms emanating from the rings.  
Figure~\ref{fig:fluct}a shows the fluctuation of the torsion angles $\phi_1$, $\phi_2$ and $\phi_3$ between the ring-units occurring in Au-h-R4, respectively. All three torsion angles fluctuate strongly around an equilibrium average of 33.9 degrees, which agrees well with the experimentally observed equilibrium value of 34 degrees. The histogram Fig.~\ref{fig:fluct}b shows that the outer torsion angles $\phi_1$ and $\phi_3$ have slightly broader distributions than $\phi_2$, which may be caused by the lower potential energy barrier at the electrodes. The average conductance over 10 ps arises as an average of strongly fluctuating instantaneous values, as illustrated in Figure~\ref{fig:fluct}c. 

In the course of the the 10 ps simulation we find 6 "near-planar" configurations of adjacent rings and 2 "near-planar" conformations of all three rings. To characterize this geometric feature we compute the average of the absolute torsion angles $\bar{\phi}=\frac{1}{3} (|\phi_1| + |\phi_2| + |\phi_3|)$, which has 4 minima (with $\bar{\phi} < 20$ deg) and 3 maxima (with $\bar{\phi} > 45$ deg) in the observation period.  As illustrated in Fig~\ref{fig:fluct}c the conductance has associated minima where $\bar{\phi}$ is maximal, e. g. at 2.2, 4.3, and 9.4 ps. Correspondingly the highest conductance values are obtained, if $\bar{\phi}$ is small, e. g. at 4.6, 5.5, and 9.8 ps. This effect is also illustrated in Fig.~\ref{fig:fluct}d, which gives a a higher time-resolution of the grey regions of Fig.~\ref{fig:fluct}a and ~\ref{fig:fluct}c. This analysis shows that the experimentally relevant conductance at room temperature arises as an average of thermally excited high-conductance conformations, which are forbidden at zero temperature. 

\section{Conclusions and Discussion}

Most experimental realizations of nanoscale junctions will contain some degree of structural disorder, which is difficult to assess in situ experimentally. The importance of thermal fluctuations on the local geometry was recently modeled for benzene molecular junctions in good agreement with experimental observation\cite{andrews:2008, cui:2001}. Here we use a recursive layer Green's function approach for the ballistic electronic transport through several disordered nanojunctions. 

We have investigated the influence of imperfect contact geometries on the conductance by studying many different possible realizations of silver and gold nanowires. Tilting the junction electrodes up to 60 degrees reduces the conductance by 20\%, while twinning the electrodes leaves the conductance nearly unchanged. We found that the introduction of up to 50\%  surface vacancies leads to only small variates of the ballistic transport properties of silver contacts, as long as the immediate vicinity  narrowest point of the junction was not affected. This analysis was supported by a study of the effects of global conformational change in silver junctions for locally conserved junction geometries. Here we find that variations in the global cluster geometry ranging from 5-260 atoms have only a weak effect on the zero-bias conduction of junctions with locally conserved geometry. 

Extending earlier work on local fluctuations we have investigated the impact of thermally induced large-scale geometric change on the conductance of oligophenylene wires. Averaging the conductance over 500 conformations obtained from a 10 ps molecular dynamics simulation at room temperature, we find temporal conductance fluctuations of more than an order of magnitude.
The average conductance, which agrees well with the experimentally observed value\cite{wold:2002}, results from high-conductivity conformations that are sterically unfavorable at zero temperature. The underlying advances in the modeling approach thus permit the detailed characterization of disorder effects, which present in almost all experimental realizations, on the conductance of molecular wires. 

\textit{Acknowledgements}

We thank Gerd Sch\"on for helpful discussions and Fabian Pauly for providing the structural data on the oligophenylens. We thank David J. Wales and Xueguang Shao for the silver nano-cluster conformations studied in this work. We acknowledge financial support by the DFG (grant WE 1863/15-1) and the use of the computational facilities at the Computational Science Center at KIST, Seoul.

\section{Appendix}

\subsection{Layer approximation}

The key approximation of the recursive-layer Green's function approach is the division of the quasi one-dimensional device region  into ,,principal layers'' (see Fig. 1(b)) perpendicular to the current flow \cite{verges:1999}. In the layer-approximation all Hamiltonian overlap matrix elements $h_{ij}$ between atoms which are separated by more than one principal layer are set to zero. Therefore, the full Hamiltonian $H$ of the device region has a block-diagonal form, with $H_{k} = \{ h_{ij} \}$ representing the Hamiltonian of the ,,principal layer'' $k$.
\begin{equation}
H = {\begin{pmatrix} 

H_1     & H_{12}  & 0       & 0      \\ 
H_{21}  & H_2     & H_{23}  & 0      \\
0       & H_{32}  & H_3     & \ddots \\
0       & 0       & \ddots  & \ddots

\end{pmatrix} }
\end{equation}
This approximation seems natural for a number of systems, e.g. polymers such as DNA, where the matrix elements between neighboring base-pairs decay rapidly with distance. Overlap matrix elements, as well as Hamiltonian matrix elements of localized atom-centered basis sets decay very quickly with the nuclear distance. Once such a set of principal layers is chosen, we can calculate the zero-bias conductance applying a recursive Green's function method. In the limit of small bias voltages Verg\'{e}s \cite{verges:1999} showed for a rectangular lattice of sites, that the conductance can be evaluated by a recursive calculation of the ,,column'' Green's function, starting at the right lead iterating to the leftmost site column of the device region.  The knowledge of the charge flux in the contact between the left most column and the left lead is sufficient to evaluate the total current of the system (due to flux conservation). Here we generalize this scheme to a non-rectangular, all-valence electron model to permit efficient material specific conductance calculations in arbitrary junctions. Since molecular wires are "quasi one-dimensional" by construction, these systems are always naturally dividable into principal layers in the transport direction. The layer width, which interpolates between the full system and single-atom layers, is a free parameter of the approximation, which must be chosen to reflect the properties of the system. 

\subsection{Recursive layer Green's function method}

Let us in the following denote as layer $0$ and $N+1$ the entire left and right electrode fragment, respectively (see Fig. 1(b)). 
The computation of the Green's function starts at the rightmost layer of the central region of the system, containing the rightmost device layer $N$ and the semi-infinite right lead layer $N+1$. Its electronic structure is reflected in the retarded Green's function matrix
\begin{equation}
G_{N,N+1}^r = {\begin{pmatrix} ES_{N}-H_{N} & ES_{N,N+1}-H_{N,N+1}   \\ 
                            ES_{N+1,N}-H_{N+1,N}   & ES_{N+1}-H_{N+1} \end{pmatrix} }^{-1},
\end{equation}
which is more conveniently expressed in terms of the layer-self-energies $\Sigma^r$. The self-energy $\Sigma_{N}^r$ of the right contact can be computed from the retarded surface Green's function $g^r(E)$ \cite{verges:1999} as 
\begin{eqnarray}
\Sigma_{N}^r & = & (ES_{N+1,N}-H_{N+1,N}) (ES_{N+1}-H_{N+1})^{-1}   \nonumber \\
               &   &  \times (ES_{N,N+1}-H_{N,N+1})  \\
               & \approx & \tau_{N+1,N}(E) \  g_{N+1}^r (E) \   \tau_{N+1,N}^{\dagger} (E)
\label{eqn:sigma}
\end{eqnarray}
Here $\tau_{ij}$ denotes the coupling matrix of the layers $i$ and $j$. We can calculate the Green's function and self-energy of every principal layer $k$ recursively, using the relations: 
\begin{eqnarray}
g_k^r (E) &=& (ES_k- H_k - \Sigma_k^r (E))^{-1} \\
\Sigma_{k-1}^r (E) &=& \tau_{k-1,k}(E) \  g_k^r (E) \ \tau_{k-1,k}^{\dagger}(E) \\ 
& & \text{with } \ k = N, ... ,2.  \nonumber
\end{eqnarray}
In a layered system, we have chosen the interface such that the velocity operator has non-vanishing terms only for orbitals connecting the left-most electrode layer (layer 0) with orbitals in layer 1. To compute the GF occurring in equation (\ref{eqn:img}) we thus need only the retarded Green's function of the system comprising layer 0 and layer 1
\begin{equation}
\mathcal{G}^r(E) = [ES_{01}-H_{01}-\Sigma_0^r(E)-\Sigma_1^r(E)]^{-1},
\end{equation}
which is easily computed from the right- and left-lead self energies $\Sigma^r_i$ $(i=0,1)$. 

The final ingredient to the calculation is the computation of material-specific electrode Green's functions 
for the left and the right reservoir. In order to avoid problems that may arise from an incommensurate decription of the electrodes and the extended molecule, we compute the electrode Green's functions using the same parametrization of the Hamiltonian. We 
assume a semi-infinite fcc-lattice with the experimental lattice constant for both electrodes and 
exploit a recursive relation for the surface Green's functions of the form  $g^{-1} = \alpha  - \beta g \beta^{\dagger}$ \cite{damle:2002,guinea:1983}, where $\alpha$ and $\beta$ denote the diagonal and off-diagonal blocks of $ES - H$ in the contacts, respectively. This equation is solved iteratively until $g$ is converged, which corresponds to the materials-specific surface Green's function of a semi-infinite system. As in previous studies\cite{xie:2008a,pauly:2008b}, we compute the surface GFs for all sites in a surface layer of two atomic planes. 

It is well known that the conductance is very sensitive to interference effects that arise form small atomic displacement\cite{rodrigues:2002,rego:2003}. These effects lead to weak oscillations in the total transmission at $T=0$ which are averaged over in most quantum transport measurements at room temperature. To account for this phenomenon we average $\tau(E)$ over a small interval $[E_F - \Delta, E_F+ \Delta]$ around the Fermi energy $E_F$, whereas $\Delta = 50$~meV~$\approx 2k_BT$ to obtain a representative value of the zero-bias conductance for comparison with experiment.  

\subsection{Extended H\"uckel-Hamiltonian}

In this investigation we have employed extended H\"uckel theory (EHT), which has been widely used in molecular electronics calculations in the past \cite{samanta:1996,dalgleish:2006,stafford:2007}, for the electronic structure calculations, but other semi-empirical methods can be used as well. Excellent semi-empirical parameterizations exists for a wide array of systems, including metals, biomolecules (e.g. DNA) or semiconductors, but the limitations of this approach (e.g. regarding transferability) are also well known.
In our calculations we employ a minimal basis set \cite{davidson:1986} of non-orthogonal Slater type orbitals $| i \rangle $, including s-, p- and d-wave functions for each atom. In this representation the matrix diagonal elements $h_{ii}$ of the Hamiltonian $H$ are approximated by the corresponding orbital ionization potentials, while the off-diagonal elements $h_{ij}, i \neq j$  are chosen as $(h_{ii} + h_{jj}) \cdot s_{ij}/2$. Here $s_{ij} = \langle i|j \rangle$ denotes the overlap matrix elements of $S$. The suitable choice of the basis functions permits a fast evaluation of the integrals $\langle i| j \rangle$.
The material specific orbital ionization potentials are  extracted from experimental data or \textit{ab initio} calculations. Here we employ a well established parameter set taken from Ref. \cite{muller:2001}. This choice of the Hamiltonian parameters is derived form electronic properties, and so it is partially suited for the description of the coherent transport properties of a quantum system. In explorative calculations we found other semi-empirical parameter sets, which have been derived from structural properties and heats of formation (such as PM6), agreed less with prior theory and experiment.  

%
%

\begin{table}
\caption{Conductance of the molecular wires R1, R2, R3, and R4 without and with the principal layer approximation compared to experimental results from Ref.~\cite{wold:2002}}
\centering
\begin{tabular}[c]{|c|c|c|c|}
\hline
 molecule & $G_{\text{no layers}}$  ($10^{-3}$G$_0$) & $G_{\text{with layers}}$  ($10^{-3}$G$_0$) &  $G_{\text{exp}}$  ($10^{-3}$G$_0$) \\
\hline
 R1 & 16.10 & 14.60 &  6.40 \\
 R2 &  4.10 &  3.80 &  1.16 \\
 R3 &  0.85 &  0.80 &  0.18 \\
 R4 &  0.20 &  0.20 &  -    \\
\hline
\end{tabular}
\label{tab:opi}
\end{table}


\end{document}